\documentclass[11pt,a4paper]{article}
\usepackage{jcappub}
\usepackage{epsfig}
\usepackage{color}
\usepackage{lineno}
\title{{\it\bf SimProp}: a Simulation Code for Ultra High Energy Cosmic Ray Propagation}

\author[1,2]{R. Aloisio,}
\author[3]{D. Boncioli,}
\author[2]{A.F. Grillo,}
\author[4]{S. Petrera,}
\author[4,5]{F. Salamida}

\affiliation[1]{INAF Osservatorio Astrofisico di Arcetri, Firenze, Italy}
\affiliation[2]{INFN Laboratori Nazionali del Gran Sasso, Assergi, Italy}
\affiliation[3]{INFN and Physics Department, University of Roma Tor Vergata, Roma, Italy}
\affiliation[4]{INFN and Physics Department, University of L'Aquila, L'Aquila, Italy}
\affiliation[5]{Institut de Physique Nucl\'eaire d'Orsay (IPNO), Universit\'e Paris 11, CNRS-IN2P3, Orsay, France}

\emailAdd{aloisio@arcetri.astro.it}

\abstract{A new Monte Carlo simulation code for the propagation of Ultra High Energy Cosmic Rays is presented. The 
results of this simulation scheme are tested by comparison with results of another Monte Carlo computation as well 
as with the results obtained by directly solving the kinetic equation for the propagation of Ultra High Energy Cosmic Rays. 
A short comparison with the latest flux published by the Pierre Auger collaboration is also presented.}


\begin{document}
\maketitle

\section{Introduction}
\label{sec:intro}

The observation of Ultra High Energy Cosmic Rays (UHECR), begun in the 1960s, represents a unique 
window opened on the most energetic acceleration phenomena in the Universe. UHECR are observed at extremely 
high energies up to $3\div5 \times 10^{20}$ eV and the determination of their characteristics is of paramount importance 
in unveiling their possible astrophysical sources and/or acceleration processes. One of the key points of their study is related 
to the propagation of UHE particles in intergalactic space. The study presented here is mainly 
devoted to this analysis, outlining a novel computation scheme to treat the propagation of UHE particles.  

The propagation of UHECR from the source to the observer is mainly conditioned by the intervening astrophysical 
backgrounds, such as the Cosmic Microwave Background (CMB) and the Extragalactic Background Light (EBL). 
Experimental observations of UHECR should be always compared with theoretical expectations 
in order to firmly determine the nature of such fascinating particles and, maybe, their sources. 

Several propagation dependent features in the spectrum can be directly connected with the chemical composition of 
UHECR and/or to the distribution of their sources \cite{GZK,dip,AloBon}. Among such features, particularly important 
is the Greisen-Zatsepin-Kuzmin (GZK) suppression of the spectrum \cite{GZK}, an abrupt suppression of the observed 
proton flux due to the interaction of UHE protons with the CMB radiation field. The GZK suppression, as follows from
the original papers \cite{GZK}, refers to protons and is a consequence of the photo-pion production process 
($p+\gamma_{CMB}\to p + \pi$) suffered by these particles interacting with the CMB radiation field. The energy position of 
the GZK cut-off, as well as the flux behavior in its proximity, can be predicted theoretically with extreme accuracy \cite{BereE1/2}.

In the case of UHE nuclei the expected flux also shows a suppression at the highest energies due to the photo-disintegration 
process on the CMB and EBL fields, with the production of secondary (lighter) nuclei and nucleons: 
$A+\gamma_{CMB,EBL}\to (A-nN)+nN$, where $A$ is the atomic mass number of the nucleus. 
The energy position of the suppression in the spectrum depends on the nuclear species, mainly on its atomic mass number $A$,
and on the details of the astrophysical backgrounds \cite{NucleiAlo}. Particularly relevant is the EBL field, which fixes the energy 
of the onset of the flux suppression \cite{NucleiAlo}.

Another important quantity that, in principle, could affect the flux behavior at the highest energies is the maximum energy 
provided by the sources $E_{max}$. In a typical scenario of rigidity dependent acceleration, the maximum acceleration energy 
of nuclei is proportional to the same quantity for protons through the nucleus atomic number (charge) $Z$, 
being $E_{max}^{nucl}=Z E_{max}^{p}$. Therefore, for sufficiently low $E_{max}^p$ the UHECR flux steepening at 
the highest energies could be directly linked with the nucleus charge following, in this case, a picture analogous to the
``knee'' behavior observed in the case of galactic CR \cite{disapp}.

As a general remark it should be stressed that UHE protons propagation is affected only by the CMB field, since its 
density is almost three orders of magnitude larger than that of EBL \cite{bkg,bkg-other}. Therefore at all energies the 
proton energy losses on CMB largely dominate over those on EBL \cite{NucleiAlo}. In the case of UHE nuclei the 
photo-disintegration process on EBL is relevant because, in the Lorentz factor range $\Gamma<2\times 10^9$ 
\cite{NucleiAlo}, it has no CMB counter-part and it changes substantially the expected fluxes. Finally, the pair-production 
process of nuclei over the EBL field, as in the case of protons, is negligible because always dominated by the CMB radiation 
field.

The EBL radiation field suffers of several uncertainties mainly connected with its cosmological evolution while the CMB 
field is analytically known at any red-shift. This fact explains why the propagation features in the spectrum of protons, 
such as the GZK-cut off, are less affected by uncertainties with respect to the ones relative to nuclei.

From 1960s a flattening has been observed in the UHECR spectrum at an energy around $3\div 6 \times 10^{18}$ eV, which was 
called "the ankle". This feature may be explained in terms of the pair-production dip \cite{dip}, that, like the GZK steepening,
can be directly linked to the interaction of protons with the CMB radiation. The dip arises due to the process of pair production 
suffered by protons interacting with the CMB field $p+\gamma_{CMB}\to p + e^+ + e^-$ \cite{dip}. It is present only in the 
spectrum of protons at energies in the range $2\times 10^{18} \div 10^{19}$ eV. The pair production process arises also in 
the propagation of nuclei, although it doesn't leave any feature in the expected spectra, through the reaction 
$A+\gamma_{CMB} \to A + e^+ + e^-$ and it involves only the CMB field, being the only EBL radiation relevant at the 
highest energies where the photo-disintegration process kicks in \cite{NucleiAlo}.

If nuclei dominate the UHECR spectrum the behavior of the observed flux, the ankle, could have a different explanation that has 
been proposed by Hill and Schramm \cite{Schramm}. They interpreted the observed ankle in terms of a two-component model; 
the low energy component being either galactic or produced by the Local Supercluster. A similar model was later considered 
also in \cite{Yoshida}. In this case the ankle energy region corresponds to the transition between two different 
components.

From the experimental point of view the situation is still unclear. The HiRes experiment shows spectral features consistent with
the proton GZK suppression and the pair-production dip \cite{HiRes}. Coherently with this picture, the chemical composition 
observed by HiRes is proton dominated at all energies $E>10^{18}$ eV \cite{HiResComp}.
Recent data from Telescope Array \cite{TA} appear to confirm this framework. The situation changes if the Auger results are taken 
into account. The Auger energy spectrum \cite{Auger} shows with high statistical accuracy the two main spectral features: ankle 
and high energy suppression, but the corresponding energies are shifted with respect to the HiRes energies by about 25\%.
However this shift could originate from the different energy scales of the experiments whose systematic uncertainty is of the 
same size. The most discrepant outcome is in the possible interpretation of the mass composition from the elongation rate data 
\cite{AugerComp} which may be interpreted as a transition from  light to a heavier composition at energies $E > 4 \times 10^{18}$ eV. 
This puzzling situation, with different experiments favoring different scenarios, shows the importance 
of a systematic study of UHECR propagation in astrophysical backgrounds. 

This paper describes a new Monte Carlo (MC) simulation code, {\it SimProp}\footnote{The {\it SimProp} code here presented is 
available for the community upon request to: \textnormal{SimProp-dev@aquila.infn.it}}, developed for the propagation of 
UHE particles (protons and nuclei) through astrophysical backgrounds. In designing such new scheme, which is not 
the first in this field of research
\cite{Allard,Stecker,Taylor,Elbert95,Epele98,Bertone02,Yamamoto04,Ave05,Armengaud05,Sigl05,Harari06,Anchor07},
we have focused on a  tool which can provide a fast and reliable analysis of the predictions on the spectrum and 
chemical composition changing the background characteristics.

In its current implementation {\it SimProp} uses a simplified nuclear model and a mono-dimensional 
treatment of the propagation, i.e. particles are propagated only in red-shift from the source to detection. More complete nuclear models 
and three-dimensional effects caused by the actual source distribution and the interaction of UHE particles with 
intergalactic and/or galactic magnetic fields will be included in further developments of the code. 

In the present paper we will present a systematic comparison of the {\it SimProp} results with the results of other MC 
schemes \cite{Allard,CRPropa} and with the analytical solution of the UHECR transport equations. 
The paper is organized as follows: in section \ref{sec:propa} we discuss our theoretical treatment of the 
propagation of UHE particles in astrophysical backgrounds, in section \ref{sec:layout} we introduce the 
layout of our MC code and its input-output, in sections \ref{sec:checks} and \ref{sec:Auger} we compare the results 
of {\it SimProp} with other computation schemes and with the Auger observations on the spectrum respectively. Finally, 
conclusions and a discussion on future developments of the code take place in section \ref{sec:conclusions}. 

\section{UHE Cosmic Ray Propagation}
\label{sec:propa}

The propagation of charged particles (protons or nuclei) with energies above $10^{17}$ eV through 
astrophysical backgrounds can be suitably studied taking into account the main channels of interaction that, 
as already anticipated in the introduction, are: 

\begin{itemize}
\item{{\it protons}} - UHE protons interact only with the CMB radiation field giving rise to the two processes of pair
production and photo-pion production. We neglect their interaction on EBL as discussed in the introduction.

\item{{\it nuclei}} - UHE nuclei interact with the CMB and EBL radiation fields, suffering the process of pair
production, in which only CMB is relevant, and photo-disintegration, that involves both backgrounds. 
While the first process conserves the nuclear species, the second produces a change in the nuclear 
species, extracting nucleons from the nucleus \cite{NucleiAlo,Stecker69}.  
\end{itemize}
 
In the energy range $E\simeq 10^{18} \div 10^{19}$ eV the propagation of UHE particles is extended over 
cosmological distances with a typical path length of the order of Gpc. Therefore we should also take into 
account the adiabatic energy losses suffered by particles because of the cosmological expansion of the Universe. 

The computational scheme used to handle the propagation of charged particles in {\it SimProp} is based on the 
kinetic approach proposed in \cite{NucleiAlo}. The main ingredients of this method are the continuous energy 
loss (CEL) approximation and the assumption of an exact conservation of the particle's Lorentz factor in the 
photo-disintegration process. Under the second hypothesis, namely neglecting the nucleus recoil in the 
interaction, we can easily separate the processes that change the Lorentz factor of the particle, 
leaving unchanged the particle type (pair and photo-pion production), from the processes that conserve it, changing the 
particle type (photo-disintegration).

The CEL approximation consists in assuming that particles lose energy (i.e. change their Lorentz factor) 
continuously. In the propagation through astrophysical backgrounds the interactions of UHE particles are naturally 
affected by fluctuations, with a non-zero probability for a particle to travel without losing energy. 
In the CEL approximation such fluctuations are neglected. 

In the case of proton propagation the CEL approximation has a negligible 
effect on the pair-production process, while in the case of photo-pion production it gives a deviation only at the highest 
energies ($E\ge10^{20}$ eV) of the order of $10\%$ with respect to the flux computed taking into account the intrinsic 
stochasticity of the process \cite{dip,Kachel06}. Having this in mind, we have chosen in {\it SimProp} to handle nucleon
propagation always under the CEL hypothesis using the analytic computation scheme presented in \cite{dip}.
Moreover, in the case of nucleons we will not distinguish between protons and neutrons because \cite{book}: 
(i) the photo-pion production process is an hadronic process that is essentially the same for protons and neutrons,
(ii) the loss length of protons is always larger than the neutron decay length, apart from the extreme energies 
(few$\times 10^{20}$ eV) where they become comparable \cite{book}. Therefore in the following we will always refer 
only to protons.
 
In the case of propagation of nuclei the energy losses due to the process of pair production can be simply related to the 
corresponding quantity for protons \cite{NucleiAlo} 
\begin{equation}
\left (\frac{1}{\Gamma}\frac{d\Gamma}{dt}\right )^{e^+ e^-}_{nuclei} = 
\frac{Z^2}{A} \left (\frac{1}{\Gamma}\frac{d\Gamma}{dt}\right )^{e^+ e^-}_{nucleons}
\label{eq:ee_loss}
\end{equation}
with $A$ being the atomic mass number of the nucleus and $Z$ its atomic mass. Thus, using the results of 
\cite{dip,Kachel06}, we can use the CEL approximation also for the process of pair-production involving 
nuclei.

The change in the Lorentz factor of the propagating particles is also linked to the cosmological evolution of the 
Universe. The expansion of the Universe causes an adiabatic energy loss to the propagating particles, that is 
(by definition) a continuous process common to protons and nuclei given by
\begin{equation}
\left (\frac{1}{\Gamma}\frac{d\Gamma}{dt}\right )^{ad}=-H(z)
\label{eq:betaad}
\end{equation}
where $H(z)=H_0\sqrt{(1+z)^3\Omega_m+\Omega_{\Lambda}}$ is the Hubble parameter at redshift $z$ in a 
standard cosmology with: $H_0=71$ $\mathrm{km/s/Mpc}$, $\Omega_m=0.24$ and $\Omega_{\Lambda}=0.72$
according to WMAP data \cite{WMAP}.

Let us now discuss the process of photo-disintegration of nuclei: this interaction changes the nucleus 
kind leaving its Lorentz factor unchanged. In the kinetic approach of \cite{NucleiAlo} the process of 
photo-disintegration is treated as a decay process that simply depletes the flux of the nucleus $A$. 
Unlike the processes discussed so far, that are scarcely affected by fluctuations, the process of 
photo-disintegration could be much more affected by the stochasticity of the interaction.
Therefore we have implemented our MC scheme only on this interaction process, which is 
simulated by computing the interaction time averaged over the density of the ambient photons:
\begin{equation}
\frac{1}{ \tau_{A,i}(\Gamma)}= \frac{c}{2 \Gamma^2} \int^{\infty}_{\epsilon_{0}(A)} d\epsilon^{'} 
\sigma_{A,i}(\epsilon^{'}) \epsilon^{'} 
\int^{\infty}_{\epsilon^{'}/(2\Gamma)} d\epsilon \frac{n_{\gamma}(\epsilon)}{\epsilon^2}
\label{eq:betadisi} 
\end{equation}
with $A$ the atomic mass number and $\Gamma$ the Lorentz factor of the interacting particle, $\epsilon^{'}$ the 
energy of the background photon in the rest frame of the particle, $\epsilon_{0}(A)$ the threshold of the considered 
reaction in the rest frame of the nucleus $A$, $\sigma$ the relative cross 
section, $\epsilon$ the energy of the photon in the laboratory system and $n_{\gamma}(\epsilon)$ the density of the 
background photons per unit energy. Equation (\ref{eq:betadisi}) is written using the Blumenthal approach 
\cite{NucleiAlo,Blum} and it refers to the specific reaction channel $i$, each characterized by a branching ratio, as reported 
in table 1 and table 2 of \cite{CrossSectionPuget}. The total inverse interaction time $\tau_A(\Gamma)$
can be obtained summing over the all possible photo-disintegration channels $i$. The photo-disintegration cross section 
as well as the relative branching ratios used in this work are taken from \cite{CrossSectionPuget}.

The dominant channels of photo-disintegration are single and double nucleon emission associated to the Giant Dipole 
Resonance (GDR) \cite{CrossSectionPuget}. These processes are favored if the energy of the background photon in the 
rest frame of the nucleus is $\epsilon<30$ MeV. At higher energies in the range $30<\epsilon<150$ MeV a multi-nucleon 
emission regime takes over, while at energies $\epsilon>150$ MeV the photo-disintegration cross section rapidly goes to 
zero \cite{CrossSectionPuget}.

Given the approximations described above, the {\it SimProp} computation scheme is a one dimensional algorithm in 
which only the red-shift $z$ follows the "history" of the propagating particle. This approximation together with
the Lorentz factor conservation in the photo-disintegration process justifies  integrating over the 
photon density, as in equation (\ref{eq:betadisi}), instead of generating the background photon 
parameters from their distribution. 

In our computation scheme the atomic mass number $A$ uniquely tags the nucleus species. Following 
\cite{CrossSectionPuget} we have chosen a list of nuclei from deuterium $(A=2)$ up to iron ($A=56$) with one stable 
isotope for each atomic mass number A. This assumption is reasonable because, as discussed in \cite{NucleiAlo,Stecker},
the radioactive decay time to the line of stability is less than the one-nucleon emission photo-disintegration loss time\footnote{This 
statement is strictly correct for all nuclei species but three unstable nuclei: $^{53}Mn$, $^{26}Al$ and $^{10}Be$.}. In the case
of mass values $A=54,50,48,46,40$ and $36$ there is more than one stable isotope but the laking of cross-section data 
for all stable isotopes makes impossible the computation, in these cases we have chosen the nucleus specie with a reasonable 
determination of the photo-disintegration cross section. A posteriori (see section \ref{sec:checks}), the agreement of our results 
with more refined computations schemes that take into account the effect of radio-active decays, such as the computations 
of \cite{Allard}, gives a solid justification to our approach. 
  
Following \cite{NucleiAlo} we are not including the photo-pion production process for nuclei. This choice is motivated 
by the fact that nuclei photo-pion production is naturally suppressed because the energy of the photon in the 
nucleus rest system is $A$ times lower than for a proton of the same energy. Differences with other simulation studies 
that include the photo-pion production for nuclei are observed only at Lorentz factors $\Gamma>10^{11}$, as we will discuss 
in the next session, producing some differences over the corresponding energy in the spectra. 
However we note that at these high energies there is not enough statistics with current experimental data 
to compare different theoretical models and approximations. In future developments of the code we will come 
back to this approximation including also the tiny effects due to the photo-pion production process for nuclei. 

The average interaction time in Eq. (\ref{eq:betadisi}) refers to the present epoch, for red-shift $z=0$. Due to the expansion 
of the universe, both the background photon density and energy will evolve with red-shift. In the case of CMB this evolution 
is known analytically: the density changes as $n_{CMB}\to (1+z)^3 n_{CMB}$ and the energy as 
$\epsilon\to (1+z)\epsilon$. The case of EBL is less clear: the EBL radiation is emitted by astrophysical objects 
at present and past cosmological epochs and subsequently is modified by red-shift and dilution due to the expansion of the
Universe. The EBL energy spectrum is dominated by two peaks: one at the optical and the other at the infra-red energies, 
produced respectively by direct emission from stars and by thermal radiation from dust. At present there are only a few 
calculations of the EBL which include cosmological evolution, most notably \cite{bkg} and \cite{bkg-other}. 
In the present paper we mainly use the EBL as presented in \cite{bkg}, which is a refinement of previous calculations 
\cite{Malkan}, based on the data from the Spitzer infrared observatory and the Hubble Space Telescope deep survey. 
In \cite{bkg} the EBL  photon density is found from $0.03$ eV up to the Lyman limit $13.6$ eV for different values of the 
red-shift up to $z=6$, after which the EBL is supposed to be zero.

\section{Monte Carlo Layout}
\label{sec:layout}

The main ingredients to initiate the simulation process are: the red-shift of the source, the primary nucleus species 
and its injection energy at the source. The simulation code propagates particles in one dimension with only
red-shift determining the particle evolution, since for a given cosmology there is a one-to-one correspondence 
between z and position of the particle. Consequently the particle is propagated in steps of red-shift. 
The MC follows the initial nucleus, secondary nuclei and protons produced 
at each photo-disintegration interaction calculating their losses up to the observer, placed at red shift zero. 
As discussed in section \ref{sec:propa}, the propagation of UHE particles is based on the analytical scheme described 
in \cite{dip} for protons and in \cite{NucleiAlo} for nuclei. The latter is modified for the use in a MC approach as described 
below.

In this initial implementation the nuclear model adopted in {\it SimProp} is quite simple: following \cite{CrossSectionPuget} we fix a 
list of nuclei that can be propagated, whose photo-disintegration cross-section is given in the same paper.
Each nuclear species in the list is univocally identified by its atomic mass number $A$ with steps of $\Delta A$ =1
starting from  iron $A=56$ down to beryllium $A=9$. The unstable nuclei with $5 \le A \le 8$ are excluded from the list and, for 
masses lower than $A=9$, only helium $A=4$, tritium $A=3$ and deuterium $A=2$ are included. 

The values of the energy threshold for single or double nucleon emission are taken from \cite{CrossSectionStecker}. In the case 
of isotopes with the same mass we choose the nucleus with the minimum value of the energy threshold for the emission of one 
nucleon (neutron or proton). The energy threshold for the emission of two nucleons is chosen again as the minimum among the 
three different values for the emission of any pair of nucleons.

\begin{figure}[!htb]
\centering
\includegraphics*[scale=.4]{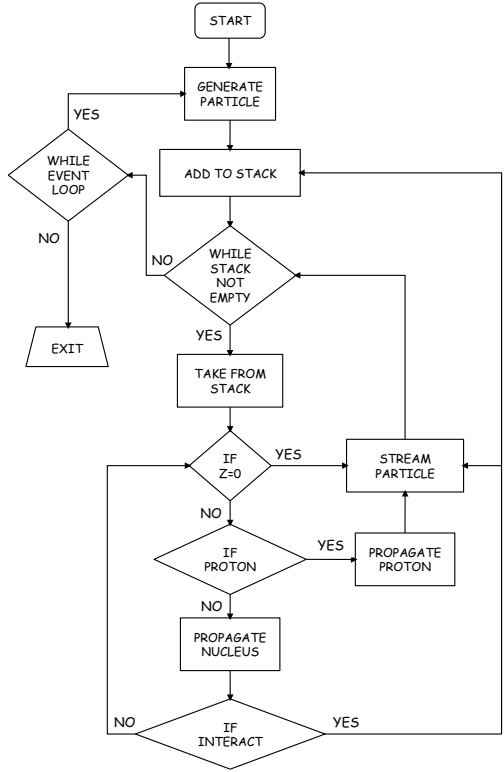}
\caption {\label{fig1}{\small{Flow chart of the simulation code {\it SimProp}.}}}
\end{figure}

Stochastic interactions must necessarily link a parent nucleus with one, or more, nuclei belonging to the list. The 
extracted nucleons ($\Delta A$) are all treated as protons, as already discussed in the previous section. 
The choice of a nuclear model based on a limited list of nuclei relies on the hypothesis that a few representative 
processes can mimic a more complex description saving the computation time. The validity of this approximation, 
already used by different authors \cite{Allard,Stecker,Taylor}, will be verified a posteriori.

The assumptions described above have an immediate consequence in the code layout, which is schematically
sketched in figure \ref{fig1}: 
\begin{itemize}
\item nuclei follow a branch of the code where both continuous and stochastic processes occur. This is done in 
the method called {\it PropagateNucleus} through several steps each one determined by the actual occurrence of a 
stochastic process. Then a change in the nuclear species and the emission of protons occur. The steps are 
iterated up to the observer at $z=0$.
\item Protons, which {\it (within the CEL approximation)} do not suffer stochastic interactions, are treated in the 
kinetic approach with a single step from their origin up to zero red-shift. This is performed in the method called 
{\it PropagateProton}.
\end{itemize}

Let us now discuss the implementation of the stochastic treatment of the nuclei propagation. 
As described above, this is done in the method {\it PropagateNucleus}. The calculations of the 
energy evolution and of the photo-disintegration life-time (Eq. (\ref{eq:betadisi})) are 
performed step by step in red-shift. The survival probability as a function of red-shift and Lorentz factor of the nucleus $A$ is:
\begin{equation}
P(\Gamma,z) = \exp \left(- \int^{z^{*}}_{z} \frac{1}{\tau_{A}(\Gamma,z^{'})}
\left|\frac{dt}{dz^{'}}\right| dz^{'} \right)  
\label{eq:prob} 
\end{equation}
where $z$ and $z^*$ are the values of the redshift of the current step (from $z^{*}$ to $z$). In the standard cosmology the 
term $|dt/dz|$ is given by
\begin{equation}
\left | \frac{dt}{dz} \right |= \frac{1}{H_0}\frac{1}{(1+z)\sqrt{\Omega_m (1+z)^3 + \Omega_{\Lambda}}}~.
\label{eq:dtdz}
\end{equation}

\begin{figure}[htb!]
\centering
\includegraphics*[width=.5\textwidth]{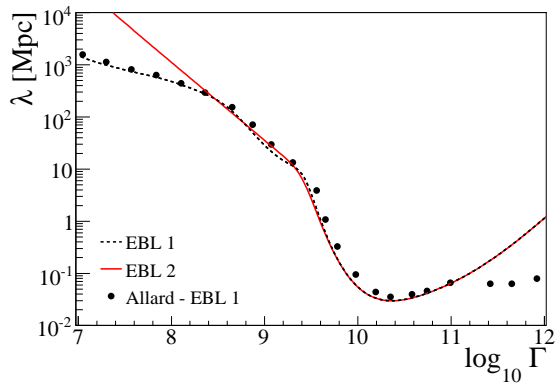}
\caption {\label{fig2}{\small{Total photo-disintegration mean free path as a function of the
Lorentz factor for iron nuclei with CMB and EBL backgrounds at red-shift $z=0$. The black dashed line 
corresponds to the path length in the case of the EBL evolution of \cite{bkg} (EBL 1), the red full line corresponds
to the broken power law approximation of \cite{bkg} (EBL 2), dots refer to the path length computed in \cite{Allard}.}}}
\end{figure}

In {\it SimProp} the intervals in red-shift which label the position of the particles have an exponentially decreasing size towards
the generation point of the nucleus. This choice assures a higher accuracy in the evolution reconstruction near the production point.
Given the probability (\ref{eq:prob}) and the energy of the nucleus at a certain step in red-shift, the MC method is applied to 
select whether the nucleus interacts and to calculate the actual interaction point within the step. The MC method is 
also applied to decide the multiplicity of the photo-disintegration. To this end we have used the (average) cross sections with the 
different branching ratios as reported in tables 1 and 2 of \cite{CrossSectionPuget}.

The {\it SimProp} program
is developed in C++. The inputs needed by the code are:
(i) the initial random seed; (ii) the number of events; (iii) the type of astrophysical background; (iv) the nucleus mass;
(v) the minimum and maximum generation energy of the nucleus; (vi) the minimum and maximum generation red-shift 
of the nucleus. The simulation code can be run injecting at the sources either a fixed primary nucleus species or any 
distribution of nuclear masses. Details about the execution performances of {\it SimProp} are given in appendix \ref{sec:Perf}.

The output of the simulation is stored in a ROOT \cite{ROOT} file recording the particles at each step of their propagation. 
The output is organized in branches containing the following information: (i) the branch of the propagation; 
(ii) the mass and the charge of the nucleus; (iii) the initial and final energy; (iv) the initial and final redshift; 
(v) the multiplicity of the interaction suffered by the current nucleus; (vi) the distance covered in the current step.
The branch number zero refers to the primary nucleus. Nuclei and protons produced by 
photo-disintegration are traced branch by branch till they reach the observer at $z=0$. 

The code is designed in such a way that any red-shift distribution of sources and any injection spectrum 
can be simulated. This is achieved drawing events from a flat distribution in the red-shift of the sources and 
of the logarithm of the injection energy. Once the event is recorded at $z=0$ the actual source/energy distribution is recovered 
through a proper weight attributed to the event. As an example, let us discuss the case of uniformly distributed 
sources in co-moving coordinates with a power law injection spectrum. In this case events should be weighted with a factor
\begin{equation}
w_z \propto \frac{1}{(1+z)\sqrt{(1+z)^3 \Omega_m+\Omega_{\Lambda}}}~,
\label{eq:z_w}
\end{equation}
with $z$ the source red-shift. In the same way to generate a power law injection spectrum, with spectral index 
$\gamma$ and generation energy $E_g$, a weight
\begin{equation}
w_E \propto E_g^{1 - \gamma}
\label{eq:E_w}
\end{equation}
has to be assigned to each event at $z=0$. 

\begin{figure}[htb!]
\centering
\includegraphics*[width=.5\textwidth]{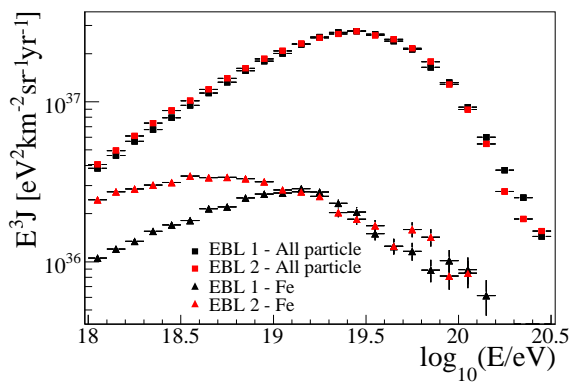}
\caption {\label{fig3}{\small{Flux of Fe (lower curves) and all particle (upper curves) in the case of pure Fe injection 
with a power law index $\gamma=2.2$ and no energy cutoff at the source. Black: calculated with the EBL given in 
\cite{bkg}, EBL 1. Red: calculated with the analytical parametrization of the EBL background, EBL 2 (see text).}}}
\end{figure}

Some of these inputs require a choice among options, depending on the 
specific needs of the user. As it is easy to understand, depending on the different options the performances of the 
simulation code could change. 

One of the input required by {\it SimProp} is connected with the EBL evolution model which is not analytically known, as 
discussed in section \ref{sec:propa}. The EBL evolution assumed in {\it SimProp} is the one given in \cite{bkg} (EBL 1)
or an analytical approximation of it based on a broken power law behavior (EBL 2); the latter choice assures a faster 
computation time. Other possible assumptions on the EBL evolution can be found in \cite{bkg-other}. These different
choices for the EBL evolution were tested with essentially the same final results: the only difference connected with this 
choice regards the execution time of the simulation. 

The total photo-disintegration path length for iron as function of the
Lorentz factor is shown in figure \ref{fig2}. The black dashed line is
calculated assuming the EBL evolution reported in \cite{bkg} (EBL 1), while
the red full line is obtained using its broken power law approximation (EBL 2) .  
The black dots represent the path length as computed in \cite{Allard}. 
The differences between (EBL 1) and the simple analytical 
approximation (EBL 2) are limited to the low energy region, where the effect of 
photo-disintegration is not relevant being the corresponding path length of the 
same order of the Universe size (at Gpc scale). The differences with respect to the results reported in
\cite{Allard} are sizeable only at the highest energies where it is relevant
the effect of the photo-pion production and the weight of those nuclei
that we have neglected in our simulation (see the discussion of
section \ref{sec:propa}).

In any case what is important to discuss here is the effect on the observed flux on Earth due to the different possible 
assumptions on the EBL evolution. In figure \ref{fig3} we plot the flux of primary iron expected on Earth 
in the case of an injection spectrum $\propto E_g^{-2.2}$ with uniformly distributed sources in co-moving coordinates.
Here and in the following all fluxes are normalized to the Auger spectrum \cite{Auger2011} above $10^{18.8}$ eV. 
The red triangles refer to the (EBL 1) while the black triangles to the simple broken power law behavior (EBL 2). Sizable 
differences in the iron spectra are recognizable only at energies below $10^{19}$ eV. In the same figure the all particle 
spectrum is shown, i.e. the sum of primary iron and all secondaries produced by photo-disintegration along the propagation 
path: in this case the effect of the different choices for the EBL is negligible.

\begin{figure}[htb!]
\centering
\includegraphics*[width=.5\textwidth]{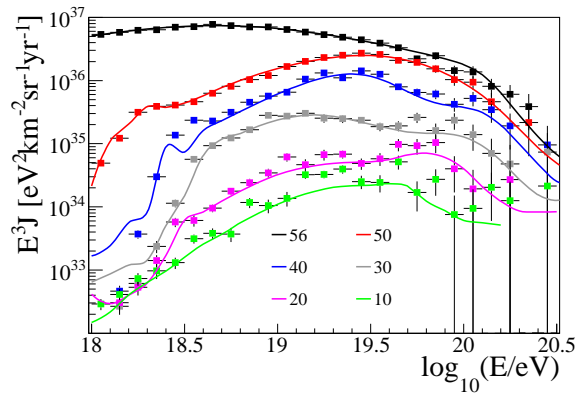}
\caption {\label{fig4}{\small{Flux of iron and secondary nuclei (A=50, 40, 30, 20, 10) at $z=0$ in the case of pure iron injection at 
the source with a power law injection index $\gamma=2.2$. Full squares correspond to the {\it SimProp} result
while continuous lines correspond to the solution of the nuclei kinetic equation of \cite{NucleiAlo}.}}}
\end{figure}

\section{Comparison with other propagation schemes}
\label{sec:checks}

In this section we discuss the comparison between the results of {\it SimProp} and other computations schemes 
based: (i) on a pure kinetic approach and (ii) on different MC schemes. In particular, since {\it SimProp} is based on 
the kinetic approach of \cite{NucleiAlo}, a comparison with the results obtained in such a scheme is of particular importance 
in order to assess the internal consistency of our MC code. To compare the {\it SimProp} results with other MC
computation schemes we have chosen the simulation by Allard et al. in \cite{Allard} and the CRPropa \cite{CRPropa} 
simulation code. Let us discuss separately the two cases.

\subsection{Kinetic Approach}
\label{check1}

In this sub-section, the spectra obtained using {\it SimProp} have been compared with those calculated
solving the kinetic equation associated to the propagation of nuclei \cite{NucleiAlo}. To pursue such comparison, a pure 
iron injection with a power law injection of the type $\propto E_g^{-\gamma}$ with $\gamma=2.2$ have been assumed. 
The sources have been assumed to be homogeneously distributed in the red-shift range $0<z<3$. In figure \ref{fig4} the 
fluxes expected at $z=0$ are shown for iron and secondary nuclei produced in the photo-disintegration chain suffered
by primary injected irons. The points refer to the {\it SimProp} results while the continuous lines to the fluxes computed in 
the kinetic approach \cite{NucleiAlo}. A good agreement between the two schemes is clearly visible in figure \ref{fig4}. 
At the highest energies the path-length of iron nuclei is very short (lower than few Mpc, see figure \ref{fig2}). Therefore, 
to achieve a good sampling in the MC simulation,  higher statistics is needed; this is the reason for larger errors bars
in the {\it SimProp} results at the highest energies and for their less good agreement with the solution of the
kinetic equation. Notice also that the simulation used for this comparison has reduced statistics respect to the other 
figures and, more importantly, here secondaries are not grouped together.

\begin{figure}[htb!]
\centering
\includegraphics*[width=.5\textwidth]{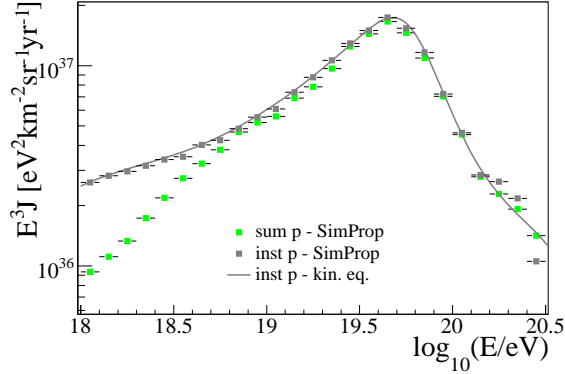}
\caption {\label{fig5}{\small{Flux of secondary protons at $z=0$ in the case of pure iron injection at 
the source with a power law injection index $\gamma=2.2$. Full squares correspond to the 
{\it SimProp} result (instanteaneous photo-disintegration approach and standard {\it SimProp} approach)
while the continuous line corresponds to the instanteaneous photo-disintegration applied to the kinetic approach 
of \cite{NucleiAlo}.}}}
\end{figure}

In the photo-disintegration chain of iron, among secondary particles, protons are also produced. As discussed in
\cite{NucleiAlo}, the flux of secondary protons can be easily computed assuming an instantaneous photo-disintegration
of the primary injected nucleus. In this case an iron nucleus once injected at the source with energy $E_g$ is 
immediately destroyed into $A=56$ nucleons each of energy $E_g/A$. At large Lorentz factors this assumption 
is well justified because the nucleus lifetime (\ref{eq:betadisi}) is much shorter than all other relevant time scales 
of the problem (see also figure \ref{fig2}). 

In figure \ref{fig5} we show the flux of secondary protons expected at $z=0$ computed in a full {\it SimProp}
simulation and assuming an instantaneous photo-disintegration of primaries. In figure \ref{fig5} we have also
computed the flux of protons obtained in {\it SimProp} forcing to zero the nuclei path-length, to mimic the 
physics of the instantaneous photo-disintegration. Figure \ref{fig5} shows the expected behavior:   
the flux obtained by a full {\it SimProp} simulation is bounded from above by the flux obtained assuming an
instantaneous photo-disintegration that coincides with the {\it SimProp} flux obtained with a null path-length 
for primaries.

In the computations presented in figure \ref{fig5} we have chosen the EBL background of \cite{bkg}, 
nevertheless the flux of secondary protons depends very little on this choice, since the EBL effect is restricted
to the Lorentz factor range $10^{8}\le \Gamma \le 2\times 10^{9}$ \cite{NucleiAlo}. 

\begin{figure}[!htb]
\centering
\begin{tabular}{ll}
\includegraphics*[width=.5\textwidth]{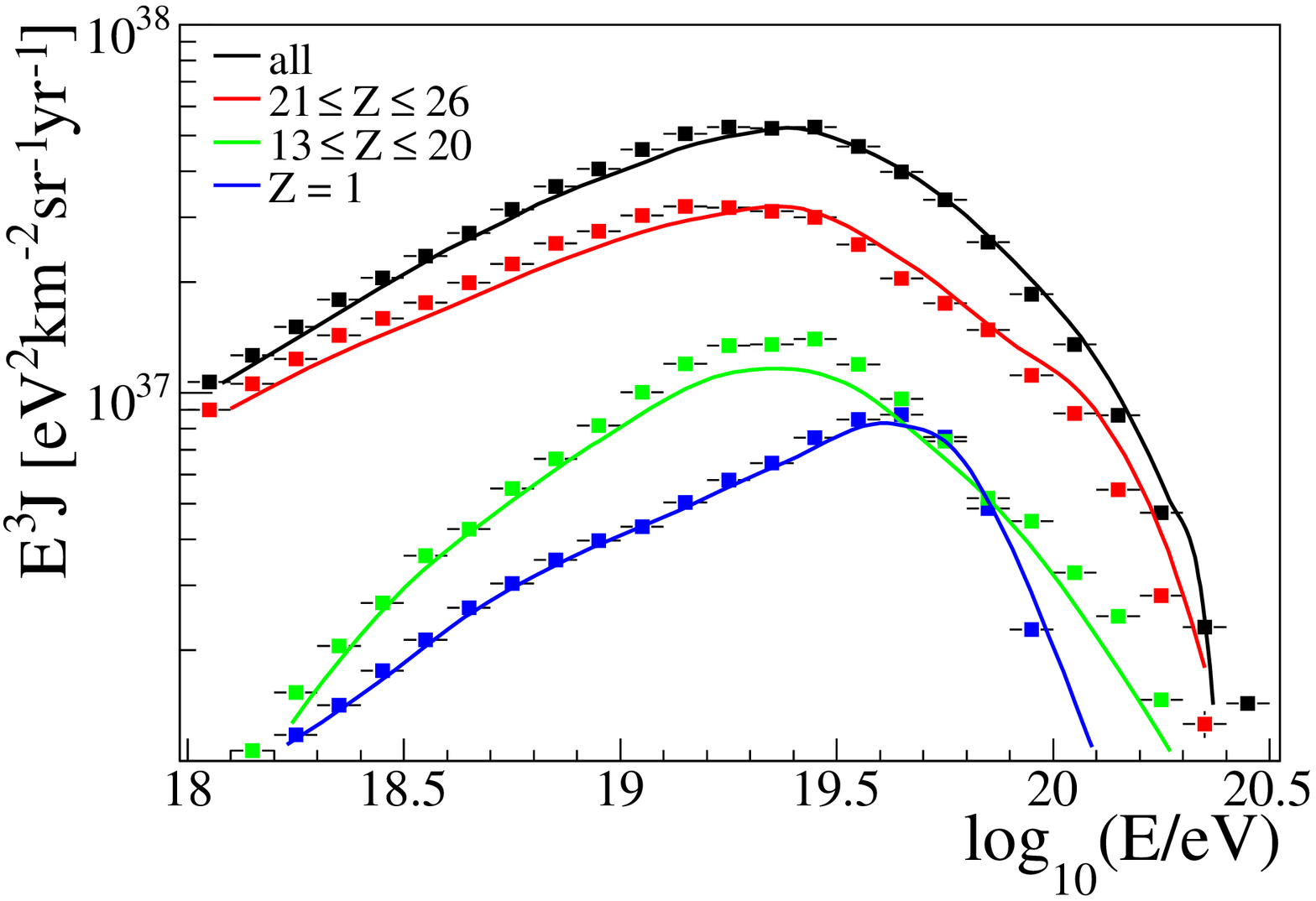}
&
\includegraphics*[width=.5\textwidth]{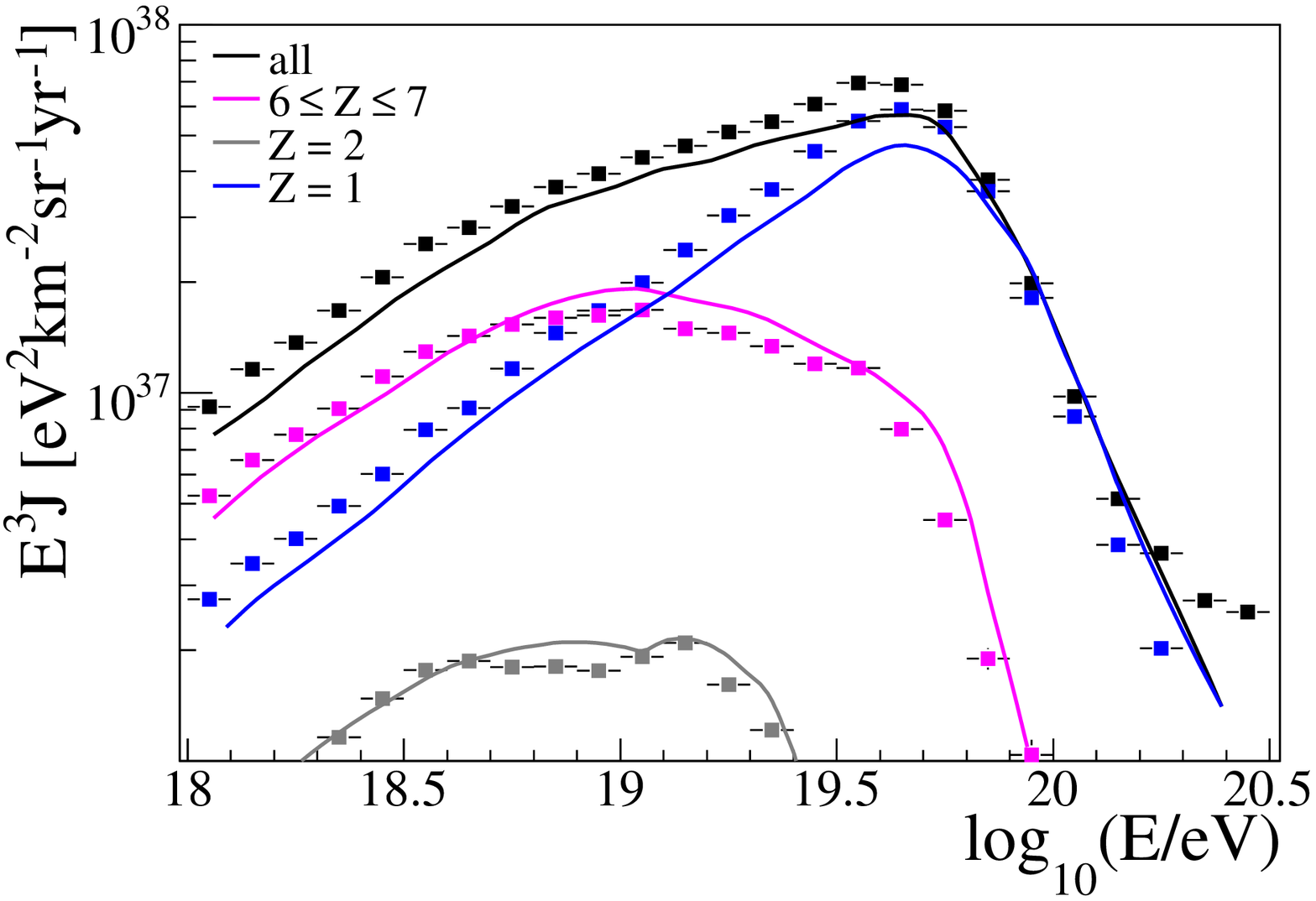}
\end{tabular}
\caption {\label{fig6}{\small{Flux at $z=0$ in the two cases of a pure iron injection with a power law injection
$\gamma=2.3$ and a maximum acceleration energy $E_g^{max}=5\times 10^{21}$ eV (left panel) and of a 
pure nitrogen injection with $\gamma=2.0$ and $E^{max}_g = 1.4\times 10^{21}$ eV (right panel).
The full squares refer to the {\it SimProp} all-particle spectrum and to the spectrum of secondary nuclei 
and protons as labeled. The continuous lines refer to the corresponding results of \cite{Allard}.}}}
\end{figure}

The agreement among the results of {\it SimProp} and those of the kinetic approach of \cite{NucleiAlo} is not 
surprising since the former is a direct derivation of the latter in which the photo-disintegration process is treated 
as a fluctuating interaction, through the MC approach discussed in section \ref{sec:propa}. Nevertheless the results 
presented in this section offer compelling evidence of the internal consistency of the computation method 
presented. 

Let us conclude this subsection discussing why it is useful to go beyond the kinetic approach. The kinetic approach 
has the important feature of being analytical: fluxes are computed mathematically solving a first principles equation \cite{NucleiAlo}. 
This means that the flux of primaries and secondaries is expressed in terms of several integrals that can be computed 
numerically, once the injection spectrum and the sources distribution are specified. In particular, the flux of secondary 
nuclei and nucleons produced in the photo-disintegration chain of a primary $A_0$ is obtained by the numerical computation 
of $A_0$ nested integrals and this computation should be repeated each time the hypothesis on sources (injection 
and distribution) are changed. This computation, while it is always feasible numerically, takes some time that can be
substantially reduced using a MC computation scheme. This follows by the fact that, as discussed in section \ref{sec:layout},
within the {\it SimProp} approach it is possible to simulate different source distributions and injection spectra without 
repeating the overall propagation of particles. In this sense the MC approach presented here, which is the minimal 
stochastic extension of the kinetic approach, provides a faster computation scheme. Finally, through the MC approach of 
{\it SimProp} one takes into account also the intrinsic stochastic nature of the photo-disintegration process which is neglected in 
the kinetic approach.

\subsection{Other MC simulations}
\label{check2}
The first code we have chosen for the comparison of {\it SimProp} with other MC approaches is the one 
presented in \cite{Allard}. We used the fluxes reported in ref. \cite{Allard}. Therefore the same injection 
conditions adopted in \cite{Allard} have been fixed for {\it SimProp}. In particular, two cases of a pure injection 
have been considered: iron nuclei with a power law injection index $\gamma=2.3$ and nitrogen nuclei with 
$\gamma=2.0$. The maximum acceleration energy is fixed to $E_g^{max}=5\times 10^{20}$ eV in the case of iron 
and to $E_g^{max}=1.4\times 10^{21}$ eV in the case of nitrogen.  The results of the comparison are 
shown in figure \ref{fig6}: left panel refers to the case of iron injection and right panel to the case of 
nitrogen.  {\it SimProp} spectra and the spectra of \cite{Allard} are normalized to Auger 
data \cite{Auger2011} above $10^{18.8}$ eV. From figure \ref{fig6} we can conclude that there is a good agreement 
among the two computations schemes in all spectrum components (primary nuclei, secondary nuclei and secondary 
protons), the largest difference being of the order of 30 \% for the flux of intermediate mass nuclei in nitrogen injection 
at $E=10^{19.75}$ eV. The results of {\it SimProp} and by Allard et al. \cite{Allard} shown in figure \ref{fig6} are obtained 
assuming the same model for the EBL background given in \cite{bkg}. 

\begin{figure}[htb!]
\centering
\includegraphics*[width=.5\textwidth]{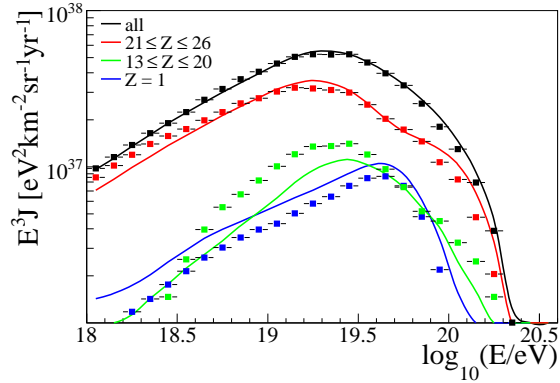}
\caption {\label{figCRP}{\small{Flux at $z=0$ in the case of a pure iron injection with a power law injection
$\gamma=2.3$ and a maximum acceleration energy $E_g^{max}=5\times 10^{21}$ eV.
The full squares refer to the {\it SimProp} all-particle spectrum and to the spectrum of secondary nuclei 
and protons as labeled. The continuous lines refer to the corresponding results of obtained with CRPropa \cite{CRPropa}.}}}
\end{figure}

The comparison between {\it SimProp} and CRPropa\footnote{We thank Simone Riggi, and CRPropa development team, for 
producing the simulation used in figure \ref{figCRP}.}~\cite{CRPropa} has been done using
the same choice of injection spectral index and maximum energy as above. The publicly available CRPropa framework - which 
was designed for the simulation of the propagation of nucleons - has been recently extended to the propagation of heavy nuclei.
The results of the comparison are given in figure \ref{figCRP}, where the spectra are normalized to Auger 
data \cite{Auger2011} above $10^{18.8}$ eV. The largest differences ($60\%$) in this comparison are found in the secondary 
flux of particles with $13 \le Z \le 20$ at $E=10^{18.75}$ eV.

The nuclear model adopted in {\it SimProp} (see section \ref{sec:layout}) is simplified with respect 
to the model used by Allard et al. in \cite{Allard} and the model used in CRPropa \cite{CRPropa}. 
The good agreement of the all-particle spectra demonstrates that a simplified scheme is effective in 
producing a reliable description of the propagated spectra, especially if we take into account the limited mass resolution 
of the experimental data.

\section{Comparison with the Auger Spectrum}
\label{sec:Auger}

In this section we compare the spectrum obtained with {\it SimProp} with the latest results of Auger \cite{Auger2011}. 
This comparison has only illustrative purposes, to show the capabilities of our computation scheme. We do not want 
here to develop a systematic study of the Auger observations in terms of spectra, which is outside the scope of this paper.

To this effect we refer to the spectrum obtained with the simulation code described in \cite{Allard} and calculated for comparison 
with Auger data presented in the ICRC 2009 \cite{ICRC09Spec}: therefore we restrict the choice 
of the injection spectral index to $\gamma=2.4$ as in this analysis \footnote{For these comparisons, we always generated  
$400000$ events in four redshift ranges: $0.0-0.02, 0.02-0.2, 0.2-1.0, 1.0-3.0$, using EBL 1.}.

\begin{figure}[!htb]
\centering
\begin{tabular}{ll}
\includegraphics*[width=.5\textwidth]{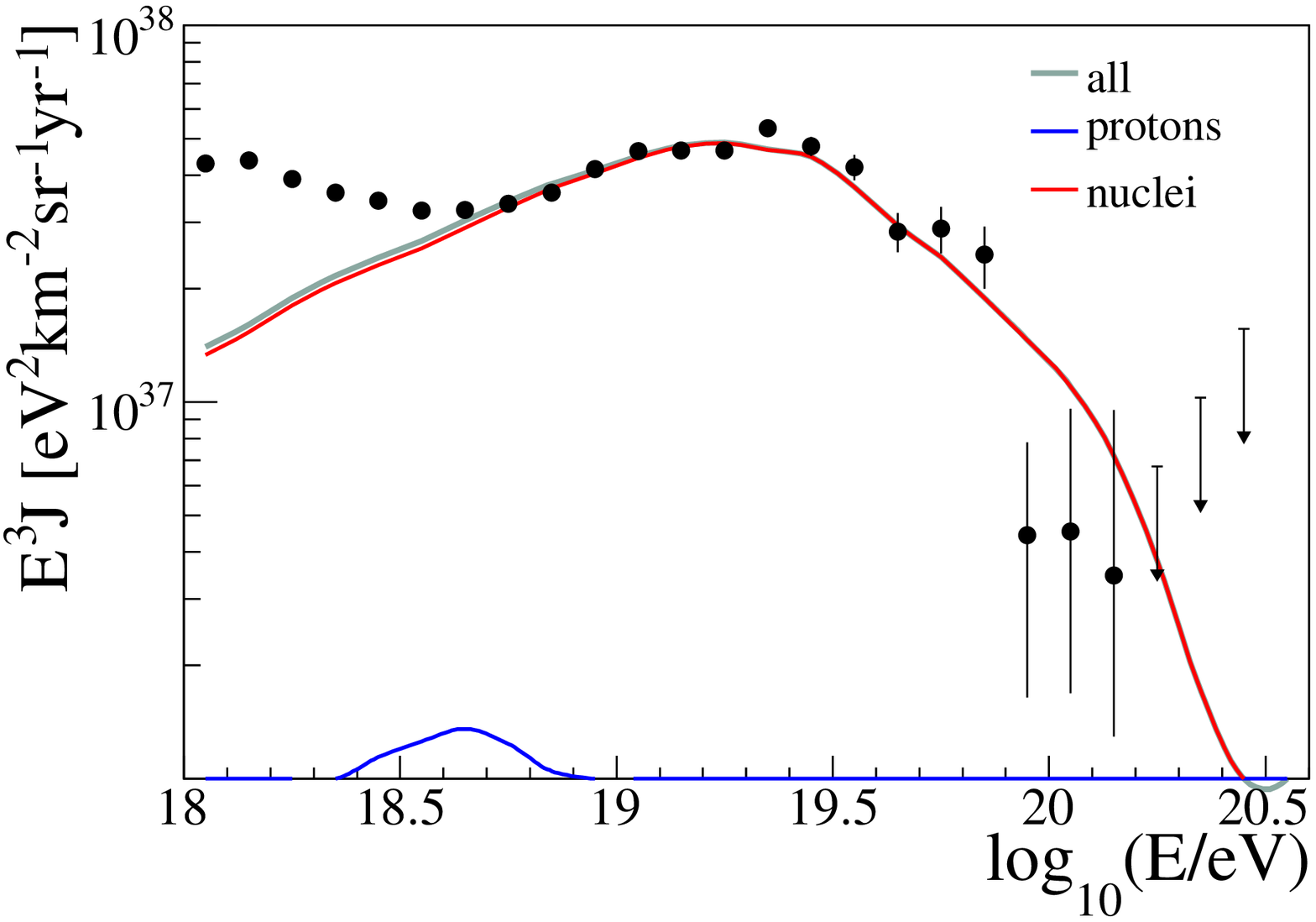}
&
\includegraphics*[width=.5\textwidth]{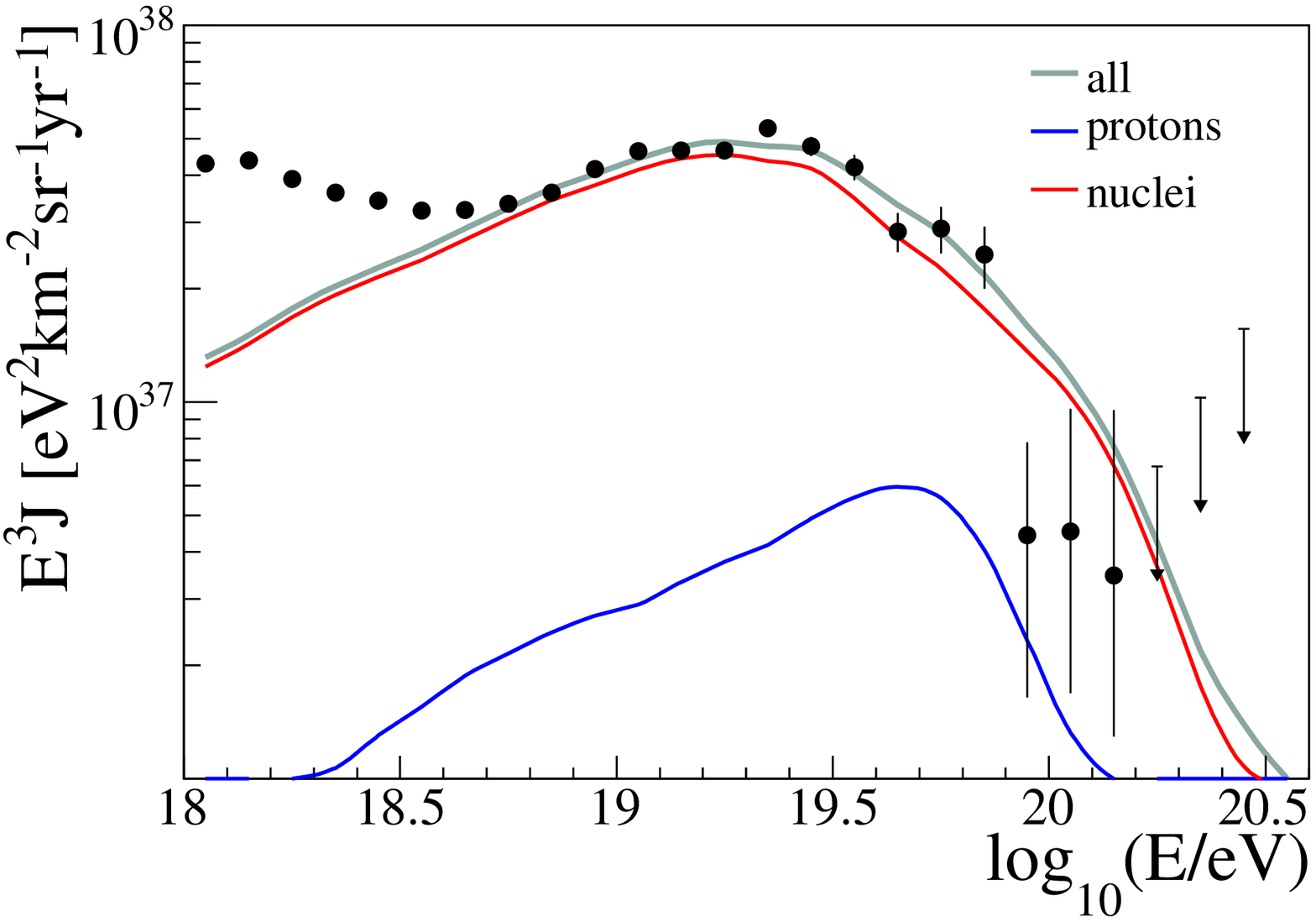}
\end{tabular}
\caption {\label{fig7}{\small {Flux of Fe nuclei 
injected with a power law with $\gamma=2.4$ and 
$E^{max}_g = 3\times 10^{20}$ eV for iron (left panel) and $E_g^{max}=\infty$ (right panel). 
The red line refers to the contributions of nuclei (summed over iron and all secondaries), the blue line 
to the contribution of secondary protons and the grey line is the all-particle spectrum. For comparison the Auger 
combined spectrum \cite{Auger2011} is shown with black circles.}}}
\end{figure}

In figure \ref{fig7} we show the spectra derived for a single component of primaries (Fe) with  two different values for 
the iron maximum energy: $E_g^{max}=3\times 10^{20}$ eV and $E_g^{max}=\infty$. The sources are assumed to be 
uniformly distributed in co-moving coordinates, and the all-particle spectra are normalized to Auger events above $10^{18.8}$ eV.

The contributions of nuclei lighter than iron to the all particle spectrum are due to the effect of photo-disintegration, 
that provides at $z=0$ a mixture of all secondary protons and nuclei with $A\leq 56$ in the photo-disintegration 
chain of iron. As expected, there is remarkable difference in the proton fraction at $z=0$ 
depending on the iron maximum acceleration energy.

From figure \ref{fig7}, we can observe that the different choices of the iron 
maximum acceleration energy have little impact on the all-particle spectra, because of the spectral cut-off 
due to photo-disintegration. However, the proton fraction at high energy is very different in the two cases 
implying some effects of this parameter on the observed elongation rate.

\begin{figure}[!htb]
\centering
\begin{tabular}{ll}
\includegraphics*[width=.5\textwidth]{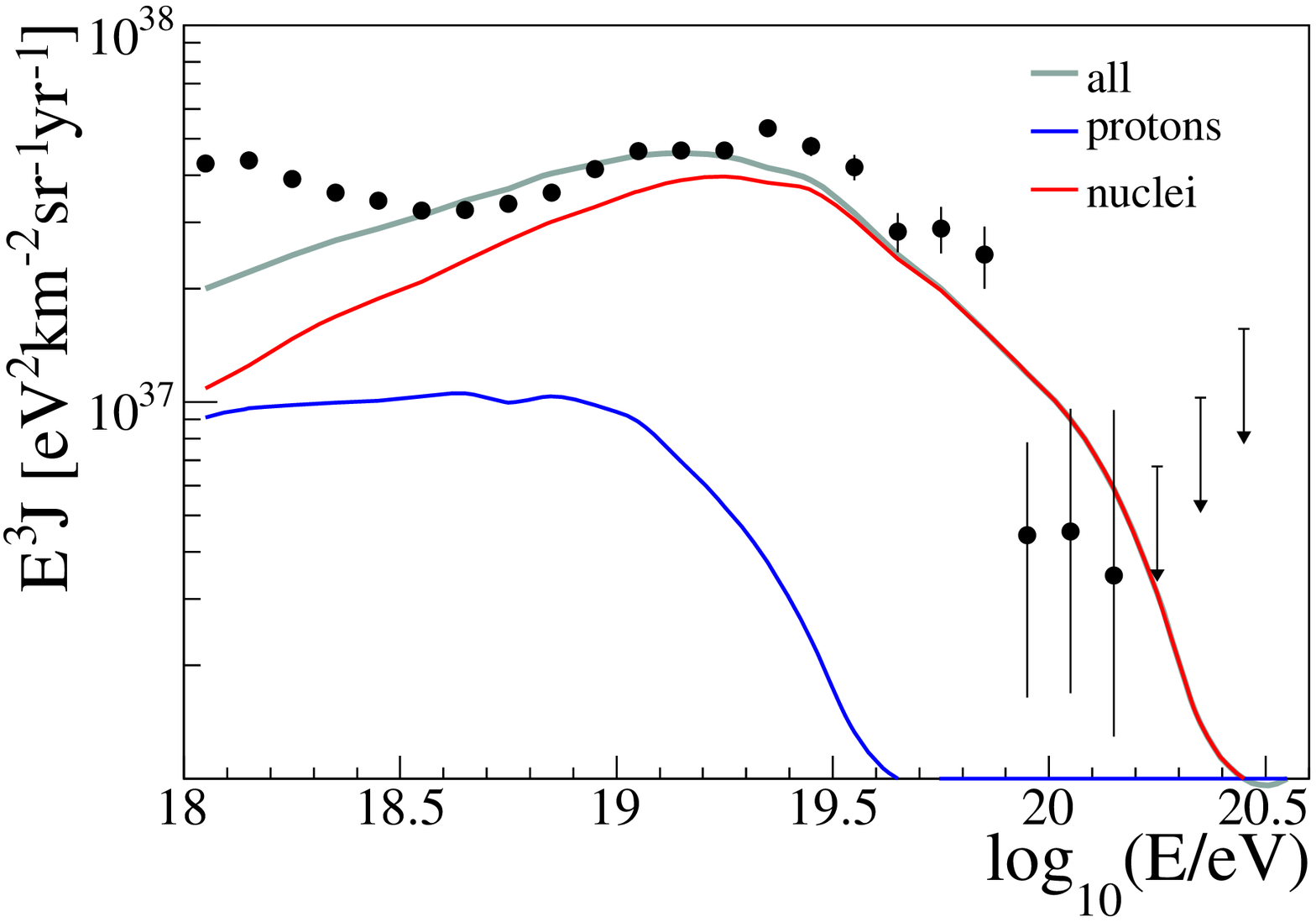}
&
\includegraphics*[width=.5\textwidth]{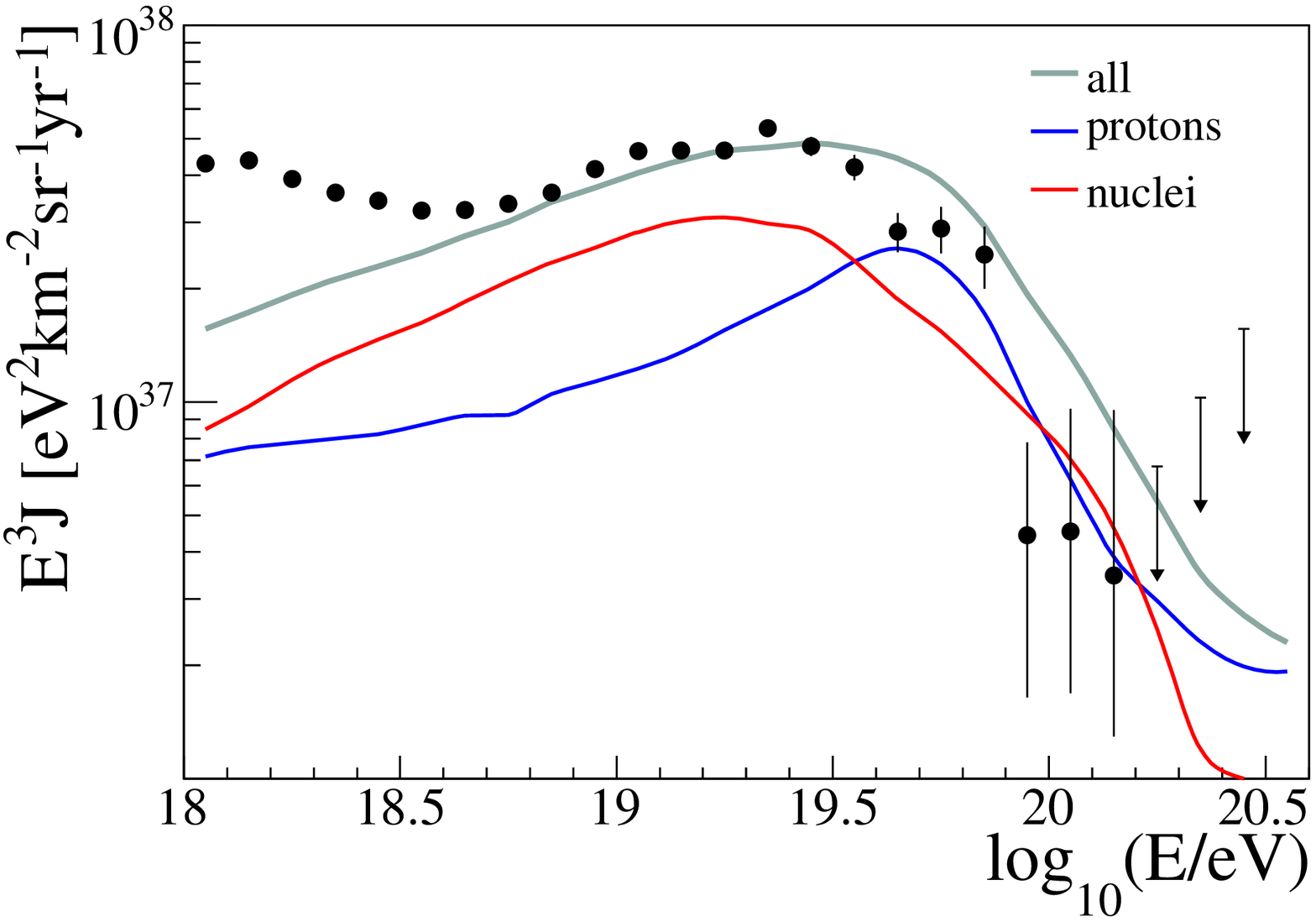}
\end{tabular}
\caption {\label{fig8}{\small{As in figure \ref{fig7} with an injection of $50\%$ Fe nuclei and $50\%$ protons, injected 
with $\gamma=2.4$ and $E^{max}_g = 3\times 10^{20}$ eV (left panel), $E_g^{max}=\infty$ (right panel).}}}
\end{figure}

In figure \ref{fig8} we plot the fluxes computed with {\it SimProp} in the case of $50\%$ injection of protons and iron nuclei, 
and a rigidity dependent energy cut-off (left panel), that is $E^{max}_g(Z)=E^{max}_g(Fe)\frac{Z}{26}$.

\section{Discussion and Future Development}
\label{sec:conclusions}

In this paper we have presented a new simulation code, {\it SimProp}, to simulate the propagation of UHE particles 
in astrophysical backgrounds. The code is based on the analytical scheme of \cite{NucleiAlo}, modified to take 
into account possible fluctuations in the photo-disintegration process of nuclei. Spectra obtained with {\it SimProp} have 
been successfully checked with spectra obtained in the kinetic approach of \cite{NucleiAlo} and with the MC 
simulation codes of \cite{Allard} and \cite{CRPropa}.

The approximations presently used in {\it SimProp} are mainly related to the nuclear model, which is based on a fixed list 
of nuclei starting from iron  down to deuterium. We have neglected here the effects due to radioactive decays of nuclei and 
to the process of photo-pion production for nuclei. In future works we will refine {\it SimProp} by including also these effects
and comparing it with other existing simulation codes. Nevertheless, as discussed in section \ref{sec:checks}, the good
agreement among fluxes obtained with {\it SimProp} and with other MC approaches  shows the tiny dependence of the results 
on the nuclear model assumed.

The performances of {\it SimProp} depend on the EBL background assumed: choosing a simple analytical approximation
for this background (parameterization suggested in \cite{bkg}) the execution time is strongly reduced, as it is shown in table 
\ref{tab.bkg} in appendix \ref{sec:Perf}. In this case {\it SimProp} is a very convenient tool to produce enough statistics by 
obtaining spectra and composition observables very quickly. Even using the complete EBL model suggested in \cite{bkg}, 
the execution time of {\it SimProp} is reasonable enabling a fast analysis of the results (see appendix \ref{sec:Perf}). 

Future developments of {\it SimProp} are planned. As a first enhancement, the introduction of the photo-pion production 
process for nuclei will be taken into account. Three-dimensional effects caused by the granularity of the actual 
source distribution and the effects of magnetic fields in the propagation of nuclei are not presently included. The results 
presented here are all obtained assuming a uniform distribution of sources in co-moving coordinates, a case which gives a flux 
independent of the magnetic field \cite{theorem}. We will also improve {\it SimProp} to carry on a systematic study of the effects 
of a sparse distribution of sources and of galactic as well as (possible) extra-galactic magnetic fields. 

Finally, in the present version of {\it SimProp} the treatment of secondary photons and neutrinos produced by the 
propagation of protons and nuclei is not implemented. It has to be noticed, however, that the main ingredients to 
determine the fluxes of such secondaries are already present in the outputs of our code, being just the production 
energy and red-shift of protons. Therefore we will soon consider also this extension of our simulation code. 

\appendix
\section{Performances}\label{sec:Perf}

In this section the averaged execution times per event (seconds) are
reported for different cases. The tests have been performed using a
virtual machine QEMU Virtual CPU at 2.27GHz with 2 GB RAM.
The reults are shown in the following tables.

The main effect on the performances of SimProp is given by the energy
of the simulated events while the influence of the redshift range is
scarcely affecting the computing times (see table~\ref{tab.zeta}).

In table~\ref{tab.nuc} it is shown the mean execution times for
different primary masses at injection. The redshift range does
not influence too much the performances of {\it SimProp}, while there is an
increase of the computation time with the primary nuclear type.

The last test was performed to compare the performance of SimProp
depending on the parametrization of the background. Two different
parameterization have been used and described in
section~\ref{sec:propa}. The comparison in table~\ref{tab.bkg} shows
an increase of the computation time of about 40 times going from EBL 2
to EBL 1. Also for the simulation of background EBL 1 the simulation
time weakly depends on the redshift range.

\begin{table}[h!]
\centering
\begin{tabular}{|l|l|l|l|}
\hline
\multicolumn{1}{|c|}{} &
\multicolumn{1}{c|}{\textbf{$0<z<0.2$}} &
\multicolumn{1}{c|}{\textbf{$0.2<z<1$}} &
\multicolumn{1}{c|}{\textbf{$1<z<3$}}  \\
\hline
\hline
$16<\log_{10}(E/{\mathrm{eV}})<18$     & 0.007   & 0.009   & 0.007 \\
\hline
$18<\log_{10}(E/{\mathrm{eV}})<20$     & 0.38 & 0.95 & 0.96 \\
\hline
$20<\log_{10}(E/{\mathrm{eV}})<21.5$   & 2.03   & 1.83   & 1.48\\ 
\hline
\end{tabular}
\caption{\label{tab.zeta}{\small{Execution times (seconds) per event (Fe injection) for different energy 
ranges of the primary and redshift ranges of the distribution of sources.}}}
\end{table}
\begin{table}[h!]
\centering
\begin{tabular}{|l|l|l|l|}
\hline
\multicolumn{1}{|c|}{} &
\multicolumn{1}{c|}{\textbf{$0<z<0.2$}} &
\multicolumn{1}{c|}{\textbf{$0.2<z<1$}} &
\multicolumn{1}{c|}{\textbf{$1<z<3$}} \\
\hline
\hline
$A=4$      & 0.18 & 0.19 & 0.19 \\
\hline
$A=14$     & 0.38 & 0.39 & 0.37 \\
\hline
$A=56$     & 1.1    & 1.3    & 1.1\\  
\hline
\end{tabular}
\caption{\label{tab.nuc}{\small{Execution times (seconds) per event for different species at injection corresponding to an 
energy range of the primary $17<\log_{10}(E/{\mathrm{eV}})<22.5$ and to different redshift ranges of the distribution of sources.}}}
\end{table}
\begin{table}[h!]
\centering
\begin{tabular}{|l|l|l|l|}
\hline
\multicolumn{1}{|c|}{} &
\multicolumn{1}{c|}{\textbf{$0<z<0.2$}} &
\multicolumn{1}{c|}{\textbf{$0.2<z<1$}} &
\multicolumn{1}{c|}{\textbf{$1<z<3$}} \\
\hline
\hline
EBL 1      & 41.5 & 44.6 & 44.5 \\
\hline
EBL 2      & 1.1  & 1.3  & 1.1  \\
\hline
\end{tabular}
\caption{\label{tab.bkg}{\small{Execution times (seconds) per event for different parametrizations of the EBL background distribution of 
photons (see text) corresponding to an energy range of the primary Fe $17<\log_{10}(E/{\mathrm{eV}})<22.5$ and to different 
redshift ranges of the distribution of sources.}}}
\end{table}

\end{document}